\documentclass[11pt]{article} 
\usepackage{epsfig,graphicx}
\usepackage{color,cite,jcappub}
\usepackage{hyperref, url}
\newcommand{\Fig}[1]{Fig.\ \ref{#1}}
\def\Eqn#1{Eq.\ (\ref{#1})}

\def\bea{\begin{eqnarray}}
\def\eea{\end{eqnarray}}
 \abstract{Modulated (p)reheating is thought to be an alternative mechanism for producing super-horizon curvature
 perturbations in CMB. But large non-gaussianity and iso-curvature perturbations produced by this mechanism
 rule out its acceptability as the sole process responsible for generating CMB perturbations. We explore the situation 
 where CMB perturbations are mostly generated by usual quantum fluctuations of inflaton during inflation,
 but a modulated coupling constant between inflaton and a secondary scalar affects the preheating process
 and produces some extra curvature perturbations. If the modulating scalar field is considered to be a dark matter candidate,
coupling constant between the fields has to be unnaturally
 fine tuned in order to keep the local-form non-gaussianity and the amplitude
 of iso-curvature perturbations within observational limit; otherwise parameters of the models have to be tightly constrained.
 Those constraints imply that the curvature perturbations generated by modulated preheating should be less than 15\% of 
the total observed CMB perturbations.
 On the other hand if the modulating scalar field is not a dark matter candidate, parameters of the models 
 could not be constrained, but the constraints on the maximum amount of the curvature perturbations coming 
 from modulated preheating remain valid.}
\begin{document}
\title{Constraints on variations in inflaton decay rate from modulated preheating}%
\author[a]{Arindam Mazumdar,}
\affiliation[a]{Theory Division, Saha Institute of Nuclear Physics, Kolkata-64, India}
\emailAdd{arindam.mazumdar@saha.ac.in}
\author[b]{Kamakshya Prasad Modak}
\affiliation[b]{Astroparticle Physics and Cosmology Division, Saha Institute of Nuclear Physics, Kolkata-64, India}
\emailAdd{kamakshya.modak@saha.ac.in}
\maketitle
\section{Introduction}\label{introduction}
Production of curvature perturbation during early stage of the universe is mainly described as a mechanism 
of quantum fluctuation generated at the time of inflation \cite{Linde:1981mu,Linde:1982uu}. But there is an equally possible theory of 
generating density perturbations via variations in the decay rate of inflaton field after the end of inflation 
\cite{Dvali:2003ar,Dvali:2003em,Vernizzi:2003vs,Kobayashi:2013nwa,Bernardeau:2004zz}.
There are some models in particle physics and string theories \cite{Cicoli:2012cy,Enqvist:2003uk,Kofman:2003nx}
where a non-renormalizable 
interaction term in the Lagrangian ensures that the decay rate for that channel varies depending 
on the vacuum expectation value (vev) of a spectator field in a particular Hubble patch. So in different 
disconnected Hubble patches the inflaton decays at different rates which produces curvature perturbations
on super horizon scales. In this paper we study both
semi-analytically and numerically how changes in coupling of the inflaton field 
with another secondary scalar field affects the number of $e$-foldings during preheating process. 
 Using the $\delta N$ approximation or ``separate universe'' approach~\cite{Sugiyama:2012tj,Lyth:2004gb,Lyth:2005fi} we  
determine the curvature perturbations produced in this period. 
We constrain the parameters of the Lagrangian from different
observables of cosmic microwave background (CMB).
    
There are three different observables that can put constraints on the model parameters. First
is the amplitude of curvature perturbations created by this mechanism. 
Second one is the amplitude of iso-curvature perturbations generated during modulated preheating. 
And third one is the non-gaussianity produced in this era. 

Variations in the decay rate of inflaton have been mainly studied in the context of modulated
reheating scenario~\cite{Dvali:2003ar,Dvali:2003em,Zaldarriaga:2003my,Ichikawa:2008ne,Suyama:2013rol,Choi:2012te}.
The term ``modulated preheating'' was first coined by the authors of
Ref.~\cite{Kohri:2009ac}. But in their calculations preheating was studied without taking into account
the effect of backreaction. Later the authors of Ref.~\cite{Enqvist:2012vx} have studied it extensively
considering the backreaction effect. They have found that the amplitude of curvature perturbations
varies non-trivially with the change in the coupling constant ($g$) between the inflaton field ($\phi$) and the secondary scalar
field ($\chi$). The reason behind this has been pointed out to be the non-trivial dependence of
Floquet exponent ($\mu_k$) with $g$. Here we study the variation of curvature perturbations 
with coupling constant $g$ using lattice simulation and develop an analytical tool which correctly
reproduces the order of magnitude of curvature perturbations coming from lattice simulation.
We find that the non-trivial dependence of curvature perturbations is there in
the results of lattice simulation too. But the variation in $g$ need to be tightly constrained
in order that the amplitude of curvature perturbations remain within the observational limit.
To constrain different parameters we have been guided by the philosophy that while building
any model of physics, the value of any particular parameter should not be fine tuned to produce the observational results.
Therefore if we assume some particular dependence of $g$ on the vev of a modulating scalar and a cutoff
scale, both of them can be constrained from the value of iso-curvature perturbations
and non-gaussianity. 

The paper is organized as follows. In section~\ref{dof} we discuss the general mechanism
of modulated preheating and dynamics of different fields. In section~\ref{mod_pre}
we develop the tool for semi-analytical estimation
of the curvature perturbations via $\delta N$ formulation in subsection~\ref{mod_pre_}. We also provide the lattice simulation
results of $\delta N$ and its derivatives in this subsection. In subsection~\ref{con_amp} 
we show the constraints on the variations in coupling constant coming from the
amplitude of curvature perturbations. The calculation of possible iso-curvature perturbations 
produced in this era has been shown in section~\ref{con_iso} and necessary constraints have
also been set. Amount of local form non-gaussianity yielded by this mechanism has been 
estimated in section~\ref{con_ng} and constraints on model parameters have been derived following
the latest Planck results. We finally sum up our findings and conclude in section~\ref{results}.
Necessary numerical details are provided in an appendix~\ref{numerical}. Throughout 
the paper we will use the units in which Planck mass $M_P= 1/\sqrt{G}=1$.

\section{Dynamics of fields}\label{dof}
We take the simple $m^2\phi^2$ inflaton potential for studying preheating mechanism. 
If there is no modulation on the coupling constant between inflaton $\phi$ and
a secondary field $\chi$ the potential can be written as 
\begin{eqnarray}
 V(\phi)={1\over 2}m^2\phi^{2} + {1\over 2} g^2\phi^2\chi^2 \; .
 \label{vphi}
\end{eqnarray}
After inflation $\phi$ oscillates around the minima of the 
potential. This oscillation pumps up the production of $\chi$ particle.
The amplitude of classical background of $\phi$ decreases as
$\phi = \phi_0 a^{-3/2}(t)$ where $a$ is the scale factor. Thus $\Phi(t) = \phi(t)a^{3/2}(t)$ maintains 
constant amplitude and has a form like,
\begin{eqnarray}
 \Phi(t) = \phi_0 \cos(mt) \; .
\end{eqnarray}
We assume that the $\chi$ field does not have any classical background value.
Rescaling the $\chi$ field as $X = a^{3/2}(t) \chi$, we find that
the Klein-Gordon equation of $X$ turns out to be in form of the Mathieu equation,
\begin{eqnarray}\label{Mathieu}
 \ddot{X}_k + \left( A_k + 2q\cos(2t) \right)X_k  = 0\; ,
\end{eqnarray}
where derivative has been taken in terms of $mt$, and 
\bea
A_k & = & {k^2\over a^2 m^2} + 2q\; , \nonumber \\ 
q & = & {g^2\phi_0^2 \over 4 m^2}\; .
\label{Aq}
\eea
Solution of $X_k$ has the form of $e^{\mu_k mt}$ where $\mu_k$
is the Floquet exponent and it is a function of $A_k$ and $q$.

\begin{figure*}
\begin{center}
\begin{tabular}{cc}
\epsfxsize=0.3\textwidth\epsfbox{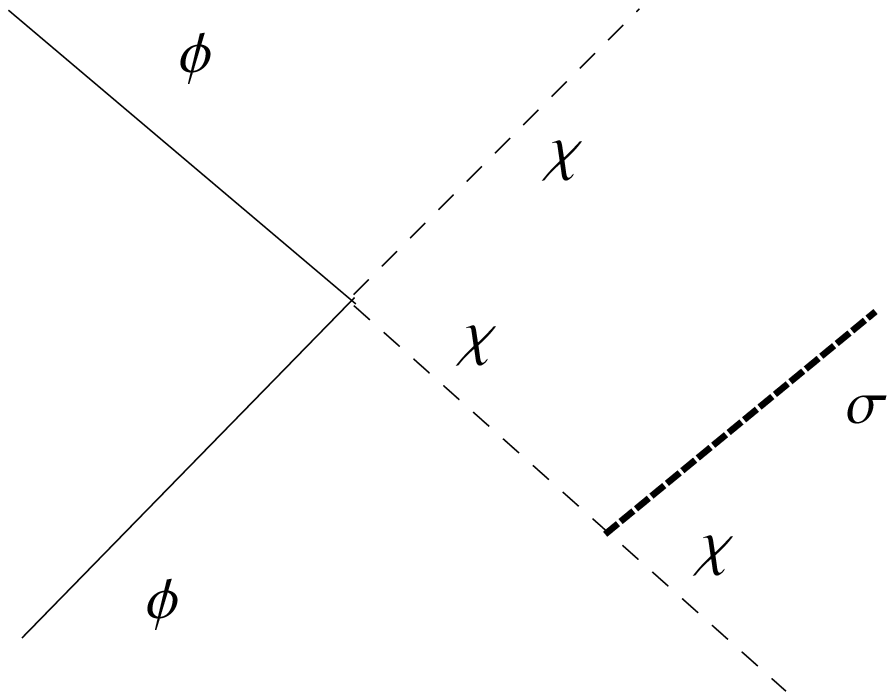}&
\; \; \; \; \; \; \; \; \; \; \; \;
\epsfxsize=0.3\textwidth\epsfbox{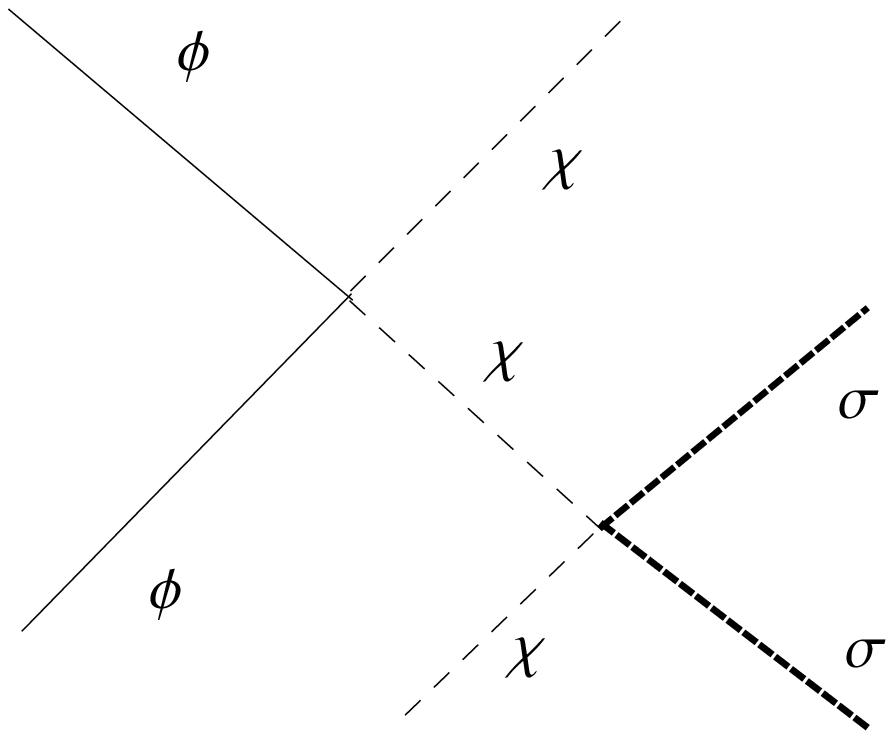}\\
(a) & \; \; \; \;\; \; \; \; \; \; \;(b)
\end{tabular}
\end{center}
\caption{(a) If there is no $Z_2$ symmetry imposed on $\sigma$, this kind of interaction will give rise to the dim-5 operator
of \Eqn{veff1}. (b) For the opposite case the dim-6 operator of \Eqn{veff2} can be generated through this interaction. }
\label{feynmann}
\end{figure*}

But if there is another scalar spectator field $\sigma$ and it is coupled to $\chi$
through some interaction term in the Lagrangian, the effective coupling
between $\phi$ and $\chi$ varies depending on the value of the vev of $\sigma$ in a particular
Hubble patch. Let us take an example 
of a particular type of dependence of $g$ on $\sigma$. In Ref.~\cite{Ackerman:2004kw} the following 
potential was assumed 
\begin{eqnarray}
 V(\phi,\sigma,\chi) = {1\over 2}m^2\phi^{2} + {1\over 2} g_0^2\phi^2\chi^2  + \xi\chi^2\sigma 
 + {\lambda\over 2}\chi^2\sigma^2 + {1\over 2}m_{\sigma}^2\sigma^2 \;, 
 \label{pot}
\end{eqnarray}
where $\xi$ and $\lambda$ are coupling constants.
{
If no particular symmetry is imposed on this potential for the field $\sigma$, 
the second term and the third term of RHS of \Eqn{pot} can generate the interaction
shown in \Fig{feynmann}-(a). This interaction can be thought of as a process to be produced by an effective 
interaction term in the Lagrangian involving two $\phi$ fields, two $\chi$ fields
and one $\sigma$ field. This effective interaction is inversely proportional to the
square of the mass ($M_{\chi}$) of the intermediate scalar field ($\chi$) in the low energy limit.
}
%
Therefore, the lowest order 
non-renormalizable interaction involving $\phi,\chi$ and $\sigma$ can be written as 
\begin{eqnarray}\label{veff1}
 g_0^2 \xi{1\over M_\chi^2}\sigma\phi^2\chi^2\; .
\end{eqnarray}
The mass of $\chi$ field comes from the vevs of $\phi$ and $\sigma$.

{
Now, if we impose a $Z_2$ symmetry on $\sigma$,
the term $\xi\chi^2\sigma$ will no longer remain in \Eqn{pot}. Therefore, the higher order interaction will be
generated from ${1\over 2} g_0^2\phi^2\chi^2$ and ${\lambda\over 2}\chi^2\sigma^2$ terms as shown by 
the diagram \Fig{feynmann}-(b).} Hence the lowest order non-renormalizable interaction would become
\begin{eqnarray}\label{veff2}
 {g_0^2 \lambda\over M_\chi^2}\sigma^2\phi^2\chi^2\; .
\end{eqnarray}
But this is not the only type of mechanism which can generate $\sigma$ dependent coupling
constant. In more realistic models fermion loops can also generate such non-renormalizable 
interactions. In string theories compactification of higher dimensions can generate modulated coupling constant\cite{Cicoli:2012cy}.
Therefore without going into the details of any particular model we keep it general and take effective coupling $g$
for \Eqn{veff1} types of cases as
\begin{eqnarray}\label{g-form-1}
 g^2=g_0^2\left(1+{\langle\sigma\rangle\over M_1}\right)\, \, ,
\end{eqnarray}
and for the case of \Eqn{veff2} it would look like 
\begin{eqnarray}\label{g-form-2}
 g^2 = g_0^2\left(1+{\langle\sigma\rangle^2\over M_2^2}\right) \; .
\end{eqnarray}
Here $M_1$ and $M_2$ are two different mass scales and $\langle\sigma\rangle$ is the classical background value 
of the field $\sigma$.

We take $m_\sigma$ to be much much smaller 
than $m$ to keep $\sigma$ field light during inflation. This will allow inflation to 
generate super-horizon perturbations of $\sigma$. If $\langle\sigma\rangle$ is of the order of $\phi$ its potential energy will
be negligible compared to ${1\over 2}m^2\phi^2$ and $\langle\sigma\rangle$
would be frozen. Therefore we can expect the value of $\langle\sigma\rangle$ to be of same order 
in the era of inflation and preheating.

The values of $M_1$ and $M_2$ determines whether the $\sigma$ field would gain sufficiently large
amplitude, comparable to that of $\chi$, at the time of preheating. Let us make an order of magnitude
estimate. It has been shown that \cite{Kofman:1997yn} to have efficient preheating $g^2$ should
be of the order of $10^5  m^2$.  The effective couplings between $\phi$ and $\sigma$ are  $g_0^2 {\langle\chi^2\rangle\over M_1}$
and $g_0^2 {\langle\chi^2\rangle\over M_2^2}$ for two above mentioned cases.
Typical value of $\langle\chi^2\rangle$ in any Hubble patch is of the order of $10^{-14}$ at the time of the start of 
preheating\cite{Kofman:1997yn}\footnote{ During inflation $\chi$ field does not get any super-horizon perturbation. It is because
for efficient preheating $g$ has to be around $10^{5}m^2$ and this value of $g$ makes $\chi$ heavy at the time of inflation
\cite{Kohri:2009ac}}.
If we take $g_0^2$ to be of the 
order of $10^5  m^2$ we need to have $M_2\approx 10^{-7}$ and $M_1 \approx 10^{-14}$ to make both the fields
$\chi$ and $\sigma$ experience parametric resonance. But when $M_2$ and $M_1$ have much higher values than these, 
only $\chi$ would undergo parametric resonance and $\sigma$ won't. After a few oscillations of $\phi$, $\langle\chi^2\rangle$ 
might be large enough to start parametric resonance in $\sigma$, but since the growth of fields under parametric 
resonance is exponential it would be too late for $\delta\sigma$ to catch up with the order of $\chi$.

The Second concern is that whether decay of 
$\chi$ into $\sigma$ would influence the parametric resonance of $\chi$. In Ref.\cite{Felder:2000hr} it had been
numerically shown that it does not. Therefore if we do not take small values of $M_1$ and $M_2$, $\sigma$ can only
show its effect on the parametric resonance of $\chi$ if it has a large finite vev $\langle\sigma\rangle$. In this
way it would modify the effective coupling constants of \Eqn{g-form-1} and \Eqn{g-form-2}. Since $m_{\sigma}$ is 
negligible compared to $H$ throughout inflation and preheating, motion of the classical background of $\sigma$ would
be over damped. Therefore we can assume that the vev $\langle\sigma\rangle$ is not very
different in inflation era and preheating era.

So, for studying the modulated preheating in different Hubble patches of the universe
we can only solve the parametric resonance of \Eqn{Mathieu} with different values of $g$.
$\langle\sigma\rangle$ is the combination of the background value of $\sigma$ at the time of 
horizon-exit of the largest mode of CMB, say ${\langle\sigma\rangle^*}$ and its super-horizon perturbations $\delta\sigma$.
\begin{eqnarray}
 \langle\sigma\rangle = {\langle\sigma\rangle^*} +\delta\sigma\; .
\end{eqnarray}
Therefore $\delta\sigma$ ensures the change in $\langle\sigma\rangle$ in different Hubble patches and 
the variation in $\langle\sigma\rangle$ is responsible for the variation in $g$.  
So Floquet exponent $\mu_k$ also varies in different regions of the universe giving rise to difference
in the time required to finish preheating. This difference in final time of preheating generates 
difference in the number of $e$-foldings $N$ in different ``seperate universes''.

\section{Amplitude of curvature perturbation}\label{mod_pre}
In this section we will derive the amplitude of curvature perturbations both semi-analytically
and numerically. Then we will compare these two results and we will also compare
our semi-analytical result with that of an earlier work. In a subsection, the
constraint coming out of the observation on the variation of $g$ will be shown. Throughout 
this section we will face no necessity for assuming any particular form of $g(\sigma)$ like \Eqn{g-form-1} and \Eqn{g-form-2},
rather we will keep $g$ as a general function of $\sigma$.

\subsection{Derivation}\label{mod_pre_}
The $\delta N$ formulation enables us to calculate super-horizon curvature
perturbation ($\zeta$) in terms of super-horizon field perturbations.
It tells us that $\delta N$, i.e the difference in the number of $e$-foldings
in different disconnected Hubble patches, is equal to the curvature perturbation
on uniform density hyper-surface.
So, for the model of our consideration we can write down, 
\bea\label{zeta}
\zeta = \delta N &=&  N_{\phi}\delta \phi + 
N_{\chi}\delta \chi + N_{\sigma}\delta \sigma  + {1\over 2}N_{\chi\chi}\delta \chi^2
+ {1\over 2}N_{\sigma\sigma}\delta \sigma^2 \; ,
\eea
where any subscript denotes derivative with respect to that field. 
Here we have assumed that during preheating inflaton does not produce any
second order perturbation\cite{Kohri:2009ac, Bond:2009xx, Mazumdar:2014haa}. 
For the $m^2\phi^2$ model in our present paper, we do not have to consider
the contribution arising from the perturbations of $\chi$, since $\chi$ does not get super-horizon
perturbation during inflation. Therefore our 
main objective would be to calculate $N_{\sigma}$ and 
$N_{ \sigma\sigma}$. 
\begin{figure*}
\centerline{\epsfxsize=0.5\textwidth\epsfbox{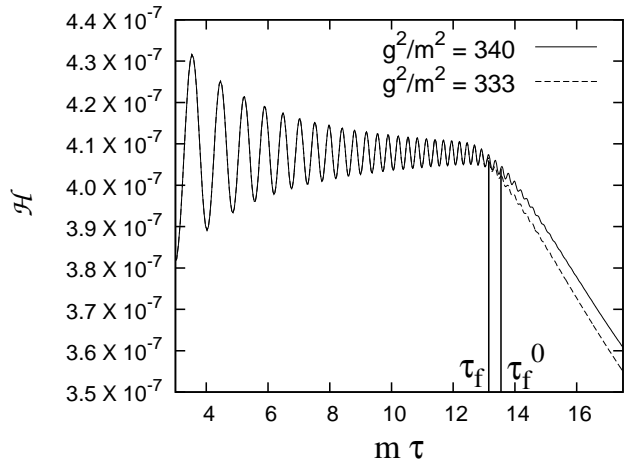}}
\caption{Trajectories of ${\cal H}$ for different $g$ values plotted using lattice simulation.}
\label{trajectories}
\end{figure*}

In $\delta N$ formulation \cite{Bartolo:2003gh,
Chambers:2008gu, Chambers:2007se} the difference in the value of the $N$ should
be taken on a uniform density hyper surface. So, for different initial conditions we have to see
the evolution of different Hubble patches up to a certain Hubble value, not up to a fixed physical time. 
Here we use the formulation 
developed in \cite{Mazumdar:2014haa} to calculate $\delta N$ in this uniform density gauge. 
We take two different values of the coupling
constant between $\phi$ and $\chi$, one is $g_0$ and another is $g$. 
Then we see the evolution of Hubble trajectories with time in two different Hubble patches for these two 
different coupling constants. We write down 
\bea
\delta N = \int_0^{t_{e}} H dt - \int_0^{t^0_e} H_0 dt \; ,
\label{deltaN1}
\eea
where $H$ is the Hubble parameter for $g$ and $H_0$ is for $g_0$ and
both the Hubble trajectories reach a uniform value $H_e$ in time $t_e$ and $t^0_e$. 
Zero in subscript or superscript denotes the trajectory for coupling constant $g_0$. 
Neglecting the gradient energy term, the Friedmann equation can be written as
\begin{eqnarray}\label{Friedmann}
 \dot H + 3 H^2 = 8 \pi V(t) \, .
\end{eqnarray}
If we change the variables to 
\begin{eqnarray}
{\cal H} = a^{3\over2} H \; ,\nonumber\\
{\cal V} = a^{3} V\; ,\nonumber\\
d\tau = dt /a^{3\over2} 
\end{eqnarray}
\Eqn{Friedmann} takes the form
\begin{eqnarray}\label{Friedmann-2}
 {d{\cal H}\over d\tau} + {3\over 2}{\cal H}^2 = 8\pi {\cal V} \, .
\end{eqnarray}
Now we break up the integration limits in the \Eqn{deltaN1}. Let us say preheating ends at a rescaled time $\tau_f$, much 
earlier than the final integration limit $\tau_e$. For the $g_0$ trajectory these times are $\tau_f^0$ and $\tau_e^0$ 
respectively. Therefore we can rewrite \Eqn{deltaN1} as
\begin{eqnarray}\label{limit-div}
 \delta N & = &\int_{0}^{\tau_f^0}\left({\cal H}(g,\tau)-{\cal H}(g_0,\tau)\right)d\tau + \int_{\tau_f}^{\tau_e}{\cal H}(g,\tau)d\tau 
  + \int_{\tau_f^0}^{\tau_f}{\cal H}(g,\tau)d\tau-\int_{\tau_f^0}^{\tau_e^0}{\cal H}(g_0,\tau)d\tau \,\,\, ,
\end{eqnarray}
{
During the period from $\tau_f$ to $\tau_e$, we need to take a particular dependence of the scale factor $a$ with the
physical time $t$. 
If $w$ is the equation of state parameter, the continuity equation,
\begin{eqnarray}\label{continuity}
 \dot\rho + 3 H(1+w) \rho = 0\, ,
\end{eqnarray}
and the definition of Hubble parameter, $H^2 = {8\pi\over 3}\rho$, 
provide us the dependence of $H$ on physical time $t$. 
After writing \Eqn{continuity} in the form of a differential equation of $H$ and 
integrating it from $t_f$ to $t$ we get
\begin{eqnarray}
 {1\over H} = {1\over H_f} +{1\over \alpha}(t-t_f)\, ,
\end{eqnarray}
where $\alpha={2\over3(1+w)}$. Further integrating over time within the same limits we find
\begin{eqnarray}\label{aform}
a(t)=\left(a_f^{1/\alpha}+ ct-ct_f\right)^\alpha \; ,
\end{eqnarray}
where $c={ H_f a_f^{1\over\alpha}\over \alpha}$ is a constant. }
This form of $a(t)$ leads us to write 
\begin{eqnarray}\label{break1}
 \int_{\tau_f}^{\tau_e}{\cal H}d\tau = -\alpha \log H_e + \alpha \log {\cal H}_f -{3\over 2}\alpha \log a_f \, .
\end{eqnarray}
We know $\log a_f = \int_{0}^{\tau_f}{\cal H}d\tau$. Again if $\tau_f<\tau_f^0$
we can write 
\begin{eqnarray}\label{break2}
 \int_{0}^{\tau_f^0}{\cal H}d\tau = \int_{0}^{\tau_f}{\cal H}d\tau +\int_{\tau_f}^{\tau_f^0}{\cal H}d\tau
\end{eqnarray}
By using \Eqn{break1} and \Eqn{break2} we can recast \Eqn{limit-div} into
\bea
\delta N &=& \underbrace{\left(1-{3\over 2}\alpha\right) \int_0^{\tau_f}({\cal H} -{\cal H}_0)d\tau}_{\delta N_1}  +
\underbrace{\left(1-{3\over 2}\alpha\right)\int_{\tau_f^0}^{\tau_f} {\cal H} d\tau}_{\delta N_2} + 
\underbrace{\alpha \log {{\cal H}_f \over {\cal H}^0_f}}_{\delta N_3} \; ,
\label{deltaN2}
\eea

Now let us evaluate the terms of \Eqn{deltaN2} one by one. First we write 
\begin{eqnarray}\label{1st-term}
\delta N_1 = \left(1-{3\over 2}\alpha\right) \int_0^{\tau_f}({\cal H} -{\cal H}_0)d\tau =
 \left(1-{3\over 2}\alpha\right) \int_0^{\tau_f}\Delta {\cal H}_i e^{\Lambda \tau}d\tau 
\end{eqnarray}
where $\Lambda$ is the Lyapunov exponent which describes the amount of separation between the 
trajectories with time. In the same equation $\Delta {\cal H}_i $ is the initial difference 
between the Hubble trajectories. In Fig.~\ref{trajectories} we see that the trajectories
with different $g$ values bend downward at different $\tau$. It means that for one of the $g$
values, preheating takes larger time than the others. In Ref~\cite{Mazumdar:2014haa} 
it was shown that for different initial
$\chi$ values, the final value of $\cal V$ changes. This shift in the final value of $\cal V$ is called 
node shift. Here for different initial $g$ values the 
node shift is negligibly small. That is 
why we do not have to use any correction term with the $\Delta {\cal H}_i$.
It is because as $g$ increases, the saturation value of $X$ decreases \cite{Kofman:1997yn} and ultimately the
final value of $g^2\Phi^2 X^2$ remains almost invariant of $g$. So the value of ${\cal V}$ at the end of preheating (${\cal V}_f$)
does not change substantially for the variation in $g$. We can understand that since the initial value of $X^2$ is negligibly 
small compared to $\Phi^2$, a change in $g$ would lead to very small value of $\Delta {\cal H}_i$. Still a large 
positive Lyapunov exponent can bifurcate the Hubble trajectories and \Eqn{1st-term} can get a large value.

Therefore we check the value of Lyapunov exponent, $\Lambda$. From the definition of Lyapunov exponent\cite{Strogatz},
we find
\bea
\Lambda (\tau) = {1\over \tau} \int_0^{\tau} \log\left(1 - 3{\cal H} + 
8\pi {\partial {\cal V} \over \partial {\cal H}}\right) d\tau' \; .
\label{lyapunov1}
\eea
from \Eqn{Friedmann-2}.
So, 
\bea
\int \Delta {\cal H}_i e^{\Lambda \tau} d\tau \approx \int \Delta {\cal H}_i \left[\exp{\int \left( - 3{\cal H} + 8\pi {\partial {\cal V} \over \partial {\cal H}} - {3\over 2}{\cal H}  \right) d\tau}\right] dt  
\eea
But ${\cal H}^2 = {8\pi \over 3} ({\cal V}+{\cal K})$, where 
${\cal K}=a^3 K$ the rescaled kinetic energy, and 
${\cal V}$ and ${\cal K}$ are of same order in preheating era. So, $2{\cal H} = 
{8\pi\over 3}\left( {\partial {\cal V} \over \partial {\cal H}}+
{\partial {\cal K} \over \partial {\cal H}}\right)$. So $3{\cal H}$ can be approximated as $8\pi {\partial {\cal V} \over \partial {\cal H}}$.
Therefore
\bea
\int \Delta {\cal H}_i e^{\Lambda \tau} d\tau \approx \int \Delta {\cal H}_i \; e^{\int - {3\over 2}{\cal H} d\tau} dt \; .
\label{lyapunov2}
\eea

\Eqn{lyapunov2} indicates that the distance between the Hubble trajectories would exponentially decrease with
time and since there is no such node shift, ultimately the total integration in \Eqn{lyapunov2} 
will produce negligible amount of $\delta N$.

 Since $g$ changes due to the change in $\sigma$, the third term in the RHS of \Eqn{deltaN2}
 can be written as
\bea
\delta N_3 =\alpha \log {{\cal H}_f \over {\cal H}^0_f}  =  
\alpha {\partial \log {\cal H}^f \over \partial \log g} {g' \over g} \delta \sigma \; .
\eea
We have already mentioned that ${\cal V}_f$ is independent of $g$. Since ${\cal K}_f$ and ${\cal V}_f$
are of the same order, ${\cal H}_f$ does not change with $g$. Therefore $\delta N_3$
would be negligible. So, the dominant contribution to $\delta N$ will arise
from $\delta N_2$ defined in \Eqn{deltaN2}. Therefore we can write $\delta N_2$ as $\delta N_{\rm MP}$, where
the subscript ``MP'' denotes modulated preheating.
Using \Eqn{break1}, we can write it as
\bea\label{deltaNexp}
\delta N_2=\delta N_{\rm MP}&= &\left(1-{3\over 2}\alpha\right)\int_{\tau_f^0}^{\tau_f} {\cal H} d\tau\nonumber \\ 
&=& \alpha_1 \left(1-{3\over 2}\alpha\right){\partial \log  H^f \over \partial \log g} {g' \over g} \delta \sigma \nonumber \\
 &=&\alpha_1\left(1-{3\over 2}\alpha\right)\left[\underbrace{{\partial \log {\cal H}^f \over \partial \log g}}_0 
+ {3\over 2} {\partial \log a_f \over \partial \log g}\right]{g' \over g} \delta \sigma \nonumber \\
&=& \left(1-{3\over 2}\alpha\right){\partial \log a_f \over \partial \log g}{g' \over g} \delta \sigma\; .
\eea
Here during the period $\tau_f$ to $\tau_f^0$ the form of $a(t)$ has been taken to be in the form of \Eqn{aform} with $\alpha$ replaced
by $\alpha_1$. 
In the last line of \Eqn{deltaNexp}, it is assumed that during preheating $\alpha_1$ is 
almost equal to ${2\over3}$.

Although we have obtained an expression for $\delta N_{\rm MP}$ in \Eqn{deltaNexp}, it is impossible to calculate the
value for a particular choice of $g$ analytically. It is because 
$a_f$ is a function of the Floquet exponent $\mu_k$\cite{Enqvist:2012vx} and  $\mu_k$ 
is not known as a function of $g$. So, we have used lattice simulation to obtain the exact values of $\delta N_{\rm MP}$ for 
different values of $g$. We have performed the simulation of preheating process for $g\over m$
from 320 to 400 with 1000 intermediate points. The simulations are run until the Hubble parameter reaches
a particular value, say $H_e$. Since $\delta N_{\rm MP}$ does not depend on the particular choice of $H_e$, any value
of it is fine as long as it is taken much after the end of preheating. The change of $N_{\rm MP}$ with $g\over m$
is shown in \Fig{Nmp}. Numerical details of the lattice simulation is given in Appendix~\ref{numerical}. 

\begin{figure*}
\centerline{\epsfxsize=1.0\textwidth\epsfbox{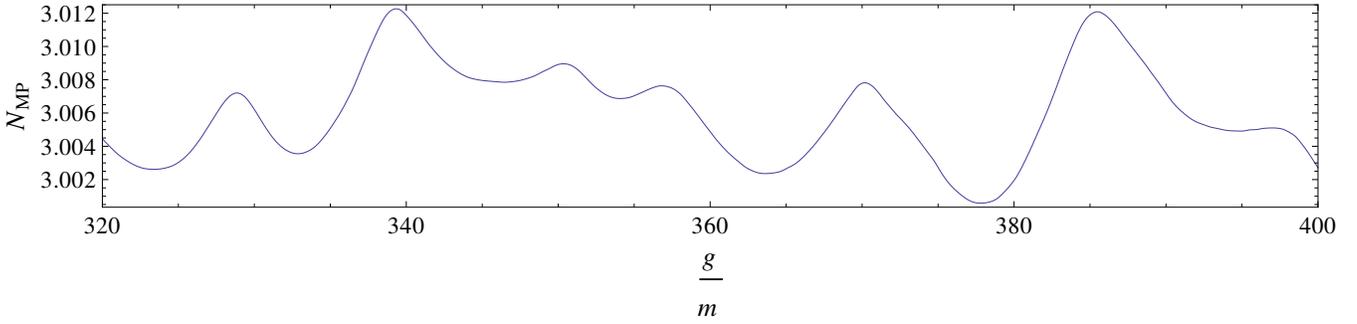}}
\caption{$N_{\rm MP}$ denotes the number of $e$-foldings during preheating period and after the end of preheating until 
Hubble parameter $H$ reaches the final value $H_e$. This plot has been produced using lattice simulation, details of which 
has been described in appendix~\ref{numerical}}
\label{Nmp}
\end{figure*}

We would like to check whether this expression of $\delta N_{\rm MP}$ gives a correct result.
So we perform an order of magnitude estimation
of the quantities of both sides in \Eqn{deltaNexp} with lattice simulation.
First we take $a_f=(1+ {3\over 2}H_i t_f)^{2\over3}$ and then from \Eqn{deltaNexp} we can write
\begin{eqnarray}\label{semi-analytic}
 {3\over 2}\delta N_{\rm MP} = \left(1-{3\over 2}\alpha\right) \delta \log (1+ {3\over 2}H_i t_f)\, .
\end{eqnarray}
From lattice simulation we have also calculated the difference between $m t_f$ for two $g\over m$ values.
For this purpose we choose two nearby $g$ values for which $\delta N_{\rm MP}$ is maximum. There is a
local maxima at ${g\over m}=340$ and a local minima at ${g\over m}=333$.
We found that for $g\over m$ values 333 and 340, the values of $m t_f$ are 83 and 90 respectively. 
If we take $\alpha$ to be $1\over 2$ then $\left(1-{3\over 2}\alpha\right) \delta \log (1+ {3\over 2}H_i t_f)$ becomes 0.018.
This is the semi-analytical result calculated using \Eqn{deltaNexp} and putting the value of $t_f$ from lattice simulation.

$\delta N_{\rm MP}$ could be calculated directly from \Fig{Nmp}.
For the values of $g\over m$ mentioned in the previous paragraph, we get
from lattice result of \Fig{Nmp} 
\begin{eqnarray}
 {3\over 2}\delta N_{\rm MP}={3\over 2}\left(N_{\rm MP}(340)-N_{\rm MP}(333)\right)=0.012\; .
\end{eqnarray} 
Since this is the same order of magnitude as the semi-analytical result derived earlier, we conclude
that \Eqn{deltaNexp} gives correct estimation.

\begin{figure*}
\centering
\includegraphics[angle=-90,scale=0.7]{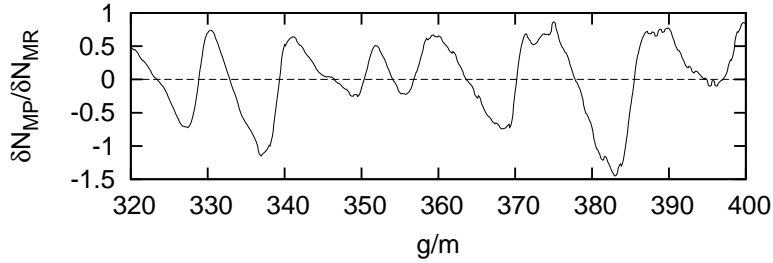}
\caption{As $g/m$ changes the ratio of the curvature perturbations generated in two different processes gets
both negative and positive values.} 
\label{amp-ratio}
\end{figure*}

Now we would like to compare our work with a similar work done earlier in Ref.\cite{Enqvist:2012vx}.
For that we need to write \Eqn{deltaNexp} in terms of curvature perturbation generated by modulated reheating.
In modulated reheating, curvature perturbation takes the following form\cite{Dvali:2003ar,Dvali:2003em}, 
\begin{eqnarray}\label{deltaNdef}
\zeta=\delta N_{\rm MR}= - {1\over 6}{\delta\Gamma\over \Gamma}=-{2\over 3}{\delta g\over g}\, ,
\end{eqnarray}
where $\Gamma$ is the decay rate from $\phi$ to $\chi$. Subscript ``MR'' denotes modulated reheating.
 Using this we can write
\begin{eqnarray}\label{compare1}
 \delta N_{\rm MP} = g {\partial N_{\rm MP}\over\partial g} {\delta g\over g} = 
 - {3\over 2} g {\partial N_{\rm MP}\over\partial g} \delta N_{\rm MR}\, .
\end{eqnarray}
$\partial N_{\rm MP}\over\partial g$ has been calculated from \Fig{Nmp} and the ratio of $ \delta N_{\rm MP}$
to $ \delta N_{\rm MR}$ has been plotted in \Fig{amp-ratio}. We can also write \Eqn{compare1} using \Eqn{semi-analytic}
and $\alpha={1\over 2}$ as 
\begin{eqnarray}\label{compare}
 \delta N_{\rm MP} = -{1\over 4} {\partial\log (1+ {3\over 2}H_i t_f)\over\partial\log g}
 \delta N_{\rm MR}\, .
\end{eqnarray}

There are some differences
between our result and that obtained in \cite{Enqvist:2012vx}. There are two reasons behind this. In the case of analytical 
calculations the authors of Ref.~\cite{Enqvist:2012vx}
did not calculate $\delta N$ in uniform density gauge. That means difference between two values $N_{\rm MP}$
for two different values of $g\over m$ was not taken
at fixed final Hubble value $H_e$.  
Rather they have fixed a final time. The final time $t$ in Eq. (3.1) of Ref.~\cite{Enqvist:2012vx}
is treated as independent of $g$. But in uniform density gauge final Hubble value should remain constant 
whereas the final time should vary.
Therefore in their estimation they got almost double the value than  the expression
shown in \Eqn{compare}. The second reason is coming from the 
fact that efficiency of lattice simulation in considering backreaction is much better than
any other semi-analytical estimation. That is why estimations of final time($t_f$) in these two methods are bound 
to be different, and consequently the earlier study \cite{Enqvist:2012vx} has shown a large ratio
of $ \delta N_{\rm MP}$ to $ \delta N_{\rm MR}$ than the lattice result shown in \Fig{amp-ratio}.

We learn from these two plots, \Fig{Nmp} and \Fig{amp-ratio}, that the curvature perturbation generated by 
modulated preheating is not large compared to that generated by modulated reheating. As far as 
the amplitude of curvature perturbation is concerned, modulated preheating is indistinguishable from
modulated reheating. 

\subsection{Constraints}\label{con_amp}
\begin{figure*}
\centerline{\epsfxsize=0.6\textwidth\epsfbox{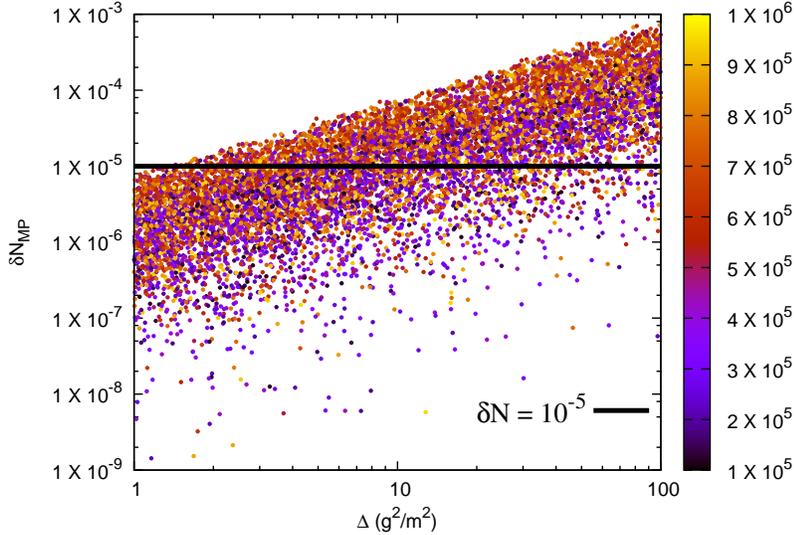}}
\caption{To keep the value of $\delta N$ under the observed bound of $10^{-5}$, variation in $g^2\over m^2$ should be about $\cal O$(1). 
Different colour represents different values of $g^2\over m^2$ around which the variation has been taken.}
\label{con-amp}
\end{figure*}
%
We have seen in \Fig{Nmp} that $N_{\rm MP}$ does not have any simple dependence on $g$ from where we could directly 
constrain its variation. Therefore 
to constrain the variation in $g$, we have taken a set of 10000 randomly different values of $g^2\over m^2$
and variation in $g^2\over m^2$( say $\Delta {g^2\over m^2}$) in \Fig{con-amp}. The set has been
generated with {log-uniform} distribution in two dimensional parameter space of $g^2\over m^2$ and $\Delta {g^2\over m^2}$.
For $g^2\over m^2$ the range was $10^{5}$ to $10^{6}$  and for $\Delta {g^2\over m^2}$ the range was 1 to 100.
From lattice simulation of \Fig{Nmp} we have generated the absolute values of $\delta N_{\rm MP}$ as
\begin{eqnarray}
 \delta N_{\rm MP} = \left| N_{\rm MP}\left({g^2\over m^2}+\Delta {g^2\over m^2}\right)- N_{\rm MP}\left({g^2\over m^2}\right)\right|
\end{eqnarray}

From \Fig{con-amp} we see that in order to keep $\delta N_{\rm MP}$ within observational bound \cite{Larson:2010gs} which
is ${\cal O}(10^{-5})$, variation in $g^2\over m^2$ has to be within $\cal O$(1). Otherwise 
to keep $\delta N_{\rm MP}$ within observational limit we have to fine tune 
$g$. That means if  $\Delta {g^2\over m^2}$ is of around $\cal O$(10) we may still have $\delta N_{\rm MP}$ under 
$10^{-5}$ but we have to choose $g^2\over m^2$ arbitrarily with an accuracy of one part in $10^6$.

Let us compare this result to the constraint from modulated reheating. From \Eqn{deltaNdef} we can say that $\delta g \over g$ has to be 
less than ${\cal O}(10^{-5})$. But in the case of modulated preheating, we have $\Delta {g^2\over m^2 }\le {\cal O}(1)$ which means
\begin{eqnarray}
{\delta g \over g}\le {1\over 2}{m^2\over g^2} \, .
\end{eqnarray}
When we are taking $g^2\over m^2$ around $10^5$ both the mechanisms are basically giving the same constraint. For higher values
of $g^2\over m^2$ this constraint becomes tighter. This is one of the main features of modulated preheating.
We have checked that in whatever range we take the values of $g^2\over m^2$ (for example from $10^6$ to $10^7$), the constraint,
$\Delta {g^2\over m^2 }\le {\cal O}(1)$ remains unchanged.

\section{Iso-curvature perturbation}\label{con_iso}
{
In this section we derive a constraint on the value of $\langle\sigma\rangle^*$.
We can think of $\sigma$ as a cold dark matter candidate if its mass is very small. In this 
case, when the Hubble parameter drops down to a value which is much less than the mass of $\sigma$, $\langle\sigma\rangle$ would
oscillate around the minima of its potential. This kind of oscillating scalar field behaves like a bosonic condensate and
can be considered as dark matter~\cite{Preskill:1982cy}.} This scenario would generate iso-curvature perturbations
whose amplitude in uniform density gauge (the same gauge in which the
$\delta N$ formulation is defined) would be \cite{Enqvist:2012vx},
\begin{eqnarray}\label{iso}
 {\cal S} \approx 2 {\delta\sigma\over {\langle\sigma\rangle^*}}\, ,
\end{eqnarray}
where the value of $\delta\sigma$ and ${\langle\sigma\rangle^*}$ should be taken at 
the time of the horizon exit of the largest mode in CMB. This form of iso-curvature
perturbation in \Eqn{iso} comes from the fact that we have assumed 
$\sigma$ to have a potential term like ${1\over 2}m_{\sigma}^2\sigma^2$. Since we are also assuming in this paper
that most of the curvature perturbations are generated by inflaton
fluctuations during inflation, the iso-curvature perturbations will be un-correlated to the 
curvature perturbations. Therefore we can say that $\delta\sigma$ is of the order of $H^*$, which
is the value of Hubble at the time of horizon exit of the largest mode in CMB.
Iso-curvature ratio is defined as \cite{Beltran:2008aa,Liddle:1993fq,Gordon:2000hv}
\begin{eqnarray}
 \alpha_{\rm iso} = {{\cal P_ S}\over {\cal P_ S}+{\cal P}_{\zeta}} = 
 {{{4\over {\langle\sigma\rangle^*}^2}\left( {H^*\over 2\pi}\right)^2}\over {{4\over {\langle\sigma\rangle^*}^2}\left( {H^*\over 2\pi}\right)^2}+{\cal P}_{\zeta}}\, ,
\end{eqnarray}
where ${\cal P_ S}$ and ${\cal P}_{\zeta}$ are the amplitudes of the power spectrums of 
iso-curvature and curvature perturbations respectively. $H^*$ has a value of $6\times10^{-6}$ for our model
where $m$ has been taken to be $10^{-6}$.
$P_{\zeta}$ is fixed at $2.2\times 10^{-9} $ from observations~\cite{Ade:2015lrj}.
Observational bound  limits 
the value of $\alpha_{\rm iso}$ to be less than 0.038~\cite{Ade:2015lrj}.
So, $\langle \sigma \rangle$ has to be taken such that $\alpha_{\rm iso}$ remains well within the observational limit.
Therefore in order to satisfy this limit we have to take,
\begin{eqnarray}\label{con-sigma}
 {\langle\sigma\rangle^*}\ge 0.2 \, .
\end{eqnarray}

{
There is a concern about this mechanism of generating the iso-curvature perturbations. It has been 
shown in Ref.~\cite{Weinberg:2004kf} that if the universe undergoes an era of local thermal equilibrium, the super-horizon 
non-adiabatic perturbations generated prior to that era would become adiabatic. After the end of preheating,
the classical background value of $\phi$ decreases and its energy is used in producing 
$\chi$ and $\phi$ particles. These particles get thermalised within some time. But there is no mechanism
which can allow the decay of $\langle\sigma\rangle$ to the constituent particles of the universe.
That is why $\langle\sigma\rangle$ will remain decoupled from the thermal bath of the universe. At a later 
time when the value of Hubble parameter drops down significantly to make $\langle\sigma\rangle$ an 
oscillating scalar field around the minima of its potential, $\langle\sigma\rangle$ will start behaving 
like a condensate of zero momentum modes and it can become a dark matter candidate  as in the
case of axion dark matter~\cite{Preskill:1982cy}. In this scenario there is no era of complete local thermal equilibrium. 
So the iso-curvature modes are expected to show up in CMB.
}

{
But if this coherent oscillation of $\langle\sigma\rangle$ leads to the evaporation of the $\sigma$ condensate,
then the local thermal equilibrium will be restored again. But that depends on many different parameters, one of 
them being the number density of $\sigma$ particles at that time. Therefore it is impossible to predict the fate
of $\langle\sigma\rangle$ condensate without taking into account the complete thermal history of the universe.
}

{
In the following section we will discuss these two possibilities. First we will put this minimum value of ${\langle\sigma\rangle^*}$ in some 
particular functional form of $g(\sigma)$ and constrain other parameters of the theory from the observational limits
of non-gaussianity. Then we will study the other case in which the constraint of \Eqn{con-sigma} is relaxed.
}

\section{Non-gaussianity}\label{con_ng}
In this section we will derive the general expression of non-gaussianity parameter $f_{\rm NL}^{\rm local}$
in modulated preheating scenario. Then we will consider one by one the models of \Eqn{g-form-2} and \Eqn{g-form-1}.  
After that we will try to put some constraints on the model parameters from observational results.
%

After picking out only the relevant terms from \Eqn{zeta}, the total curvature
perturbation can be written in Fourier space as 
\begin{samepage}
\begin{eqnarray}
 \zeta_k & = &{ N_\phi}\delta\phi_k+ { N^{\rm MP}_\sigma}\delta\sigma_k+ {1\over2}{ N^{\rm
MP}_{\sigma\sigma}}\int{d^3p\over(2\pi)^3}\delta\sigma_p\delta\sigma_{k-p}\, .
\end{eqnarray}
\end{samepage}
If the curvature perturbation generated from perturbation of $\phi$ and $\sigma$ are
uncorrelated, i.e., $\langle\zeta_k^{\phi}\zeta_k^{\sigma}\rangle=0 $, we can write
the two-point correlation function and power-spectrum as
\begin{eqnarray}\label{2-pt}
\langle\zeta_{k_1}\zeta_{k_2}\rangle & = & \langle
\zeta^\phi_{k_1}\zeta^\phi_{k_2}\rangle + \langle
\zeta^\sigma_{k_1}\zeta^\sigma_{k_2}\rangle ,\\
{\cal P}_{\zeta}(k) & = & \left(N_{\phi}^2+(N^{ MP}_{\sigma})^2\right)\left( {H^*}\over 2\pi\right)^2 \, .
\end{eqnarray}
Here we have neglected the higher order terms in $\delta \sigma$. We write the 
power spectrum of the curvature perturbations produced by modulated preheating
as 
\begin{eqnarray}
 {\cal P}_{\zeta}^{\rm MP}= (N^{\rm MP}_{\sigma})^2\left( {H^*}\over 2\pi\right)^2 \, .
\end{eqnarray} 
Defining $f_{\rm NL}$ as the ratio of bi-spectrum to the squared of power spectrum 
\cite{Komatsu:2002db,Bartolo:2004if,Liguori:2005rj} we get, 
\begin{eqnarray}\label{fnl}
f_{\rm NL}^{\rm local} & = & \left. {5\over 3}{\langle\zeta_{k_1}\zeta_{k_2}\zeta_{k_3}\rangle
 \over \langle\zeta_{k_1}\zeta_{k_2}\rangle^2+ {\rm permutations} }\right|_{k_1=k_2=k\gg k_3}\nonumber \\
 & = &  {5\over 6}(N^{ MP}_\sigma)^2 N^{ MP}_{\sigma\sigma}{{\cal P}_{\sigma}^2\over {\cal P}_{\zeta}^2}\, .
\end{eqnarray}

Unlike modulated reheating, in the case of modulated preheating we cannot write down
$N_{\sigma}$ and $N_{\sigma\sigma}$ directly as a function of $\sigma$. Rather we can only write,
\begin{eqnarray}
N_{\sigma} & =& N_g {\partial g\over\partial\sigma},\\
N_{\sigma\sigma}& =& N_g {\partial^2 g\over\partial\sigma^2} + N_{gg}\left(\partial g\over\partial\sigma\right)^2 \, .
\end{eqnarray}
Here $N_g$ and $N_{gg}$ denotes first derivative and second derivative of $N_{\rm MP}$(see \Fig{Nmp}) with respect to $g$.
Therefore for the model of \Eqn{g-form-2} we can recast \Eqn{fnl} into 
\begin{eqnarray}\label{fnl-M2}
 f_{\rm NL}^{\rm local} & = & \frac{5 {H^*}^4 {N_g}^2 g^3 {\langle\sigma\rangle^*}^2}{96 M_2^4 \pi ^4 {{\cal P}_\zeta}^2 \left(M_2^2+{\langle\sigma\rangle^*}^2\right)^3} \Biggl( M_2^2 N_{gg} g {\langle\sigma\rangle^*}^2+ N_{gg} g {\langle\sigma\rangle^*}^4 + M_2^4 {N_g} \sqrt{1+\frac{{\langle\sigma\rangle^*}^2}{M_2^2}}\Biggl)\, .
\end{eqnarray}
We have already said that in this paper that we assume curvature perturbations are mostly generated due to 
quantum fluctuations of the inflaton field during inflation. Therefore we put the value of ${\cal P}_{\zeta}$ to 
be $2.2\times 10^{-9} $ from observation~\cite{Ade:2015lrj}. 
Since $N_{\rm MP}$ behaves non-trivially with $g$ in \Fig{Nmp}, $N_g$ and $N_{gg}$ also don't have any simple dependence on 
$g$. So we take the values of $N_g$ and $N_{gg}$ from lattice results of $N_{\rm MP}$ shown in \Fig{Nmp}. By plugging in
the values of $N_g, N_{gg}, M_2$ and ${\langle\sigma\rangle^*}$ in \Eqn{fnl-M2} we can calculate the values of $f_{\rm NL}$.
\begin{table}[h!]
 \begin{center}
 \begin{tabular}{||c|c|c|c|c|c||} 
 \hline
 $M_2$ & ${\langle\sigma\rangle^*}$ & $g/m$ & $m N_g$ & $m^2N_{gg}$ & $f_{\rm NL}$ \\ [0.5ex] 
 \hline\hline
 $1\times 10^{-2}$& 0.2 &322 & -0.00048 & 0.00032 & 11.29 \\ 
 \hline
 $1\times 10^{-2}$& 0.2 &323 & 0.00009 & 0.00030 & 0.38 \\
 \hline
 $1\times 10^{-2}$& 0.2 &324 & 0.00017 & 0.00031 & 1.41 \\
 \hline
 \end{tabular}
  \end{center} 
\caption{For slightly different value of $g$, $f_{\rm NL}$ changes drastically, though $M_2$ and ${\langle\sigma\rangle^*}$
are kept fixed.}
\label{table}  \end{table}

In table \ref{table} we have demonstrated that for the same values of $M_2$ and ${\langle\sigma\rangle^*}$, $f_{\rm NL}$ can change
drastically if we make a change of order one in the value of $g\over m$. 
{
A small change in $g\over m$ changes the parameters $A_k$ and $q$ of \Eqn{Aq}. That is why the characteristic exponent ($\mu_k$)
also changes with the change in $g\over m$. So, the modes which have undergone parametric resonance earlier might not experience
resonance when $g\over m$ is changed. This leads to a sudden change in the value of $t_f$. If the values of $t_f$ obtained for 
two nearby values of $g\over m$ are very different, $N_g$ becomes large and the values of $f_{\rm NL}$ in \Eqn{fnl-M2} become
widely different. Therefore the small change in $g\over m$ can drag away the
value of $f_{\rm NL}$ from its observational limit.
Conversely, if the values of $t_f$ for different $g\over m$ are not very different, $N_g$ and $f_{\rm NL}$ both remain small.}
Any theory would be unnatural if its parameter has to be
fine tuned to $\cal O$(1) in 300 to predict observed physical quantity. Therefore to bring out the simple conclusion
from this complex dynamics we have taken help of scatter plots in the following way.

\begin{figure*}
\centerline{\epsfxsize=0.5\textwidth\epsfbox{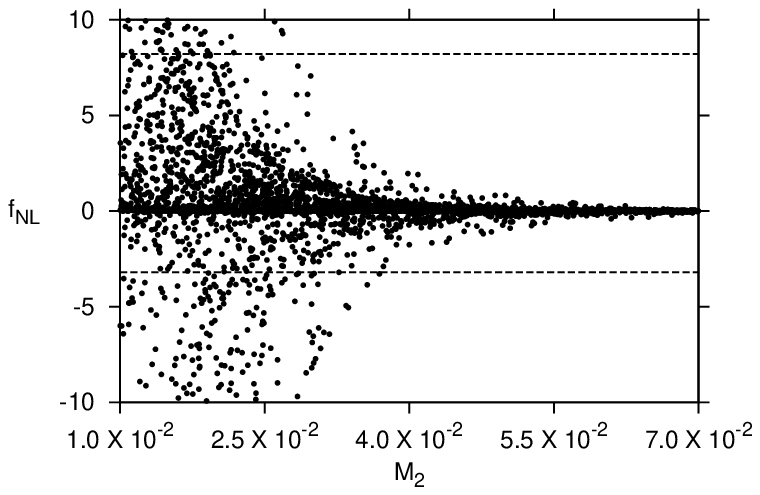}\epsfxsize=0.5\textwidth\epsfbox{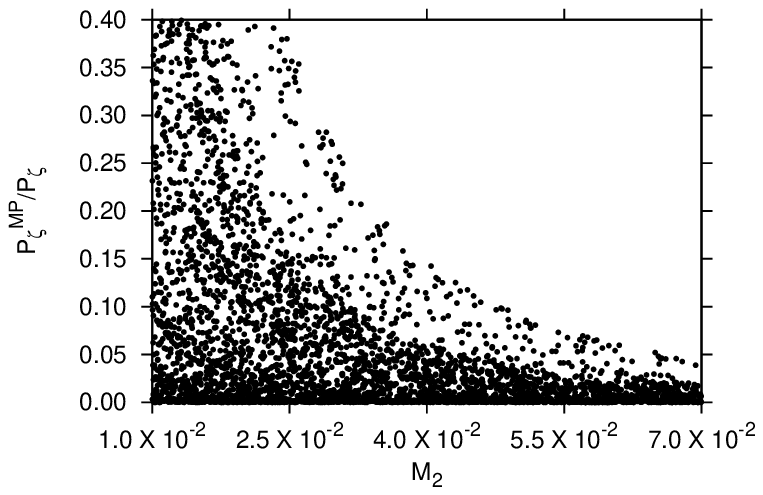}}
\caption{These plots are for $g^2=g_0^2\left(1+{\sigma^2\over M_2^2}\right)$. 
Different points in the plots represent different values of $g$ and $M_2$.
The upper and lower dashed lines indicate the observational upper and lower limit 
of $f_{\rm NL}=2.5\pm 5.7$. We see that
to keep $f_{\rm NL}$ within the observational bound for arbitrary choice of $g$ we need $M_2\ge 4\times 10^{-2}$.
${\langle\sigma\rangle^*}$ has been chosen to be 0.2 to keep amplitude of iso-curvature perturbation under 
observational constraint. We see that for that for values of $M_2$ above the threshold mentioned, ${\cal P}_{\zeta}^{\rm MP}$ to ${\cal P}_{\zeta}$ remains under 0.13. Results are in $M_P=1$ unit.}
\label{fnl-con-quad}
\end{figure*}

\begin{figure*}
\centerline{\epsfxsize=0.5\textwidth\epsfbox{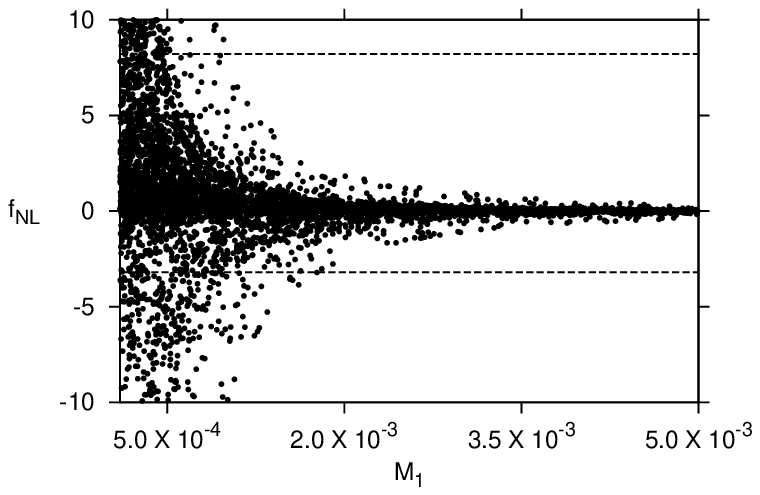}\epsfxsize=0.5\textwidth\epsfbox{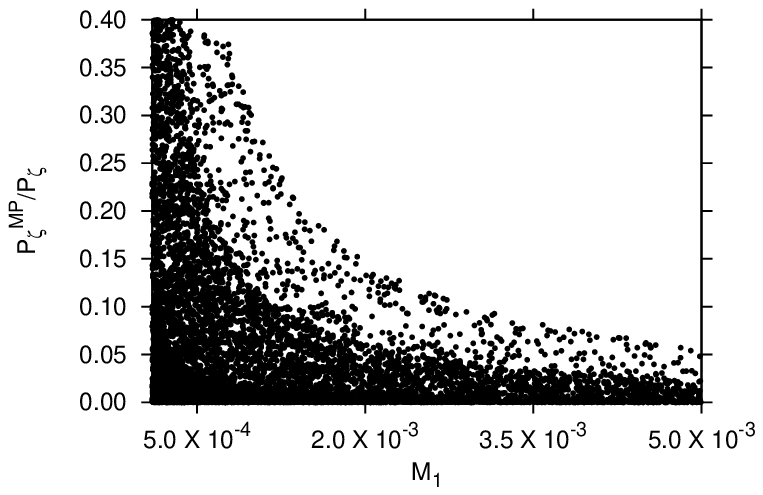}}
\caption{
These plots are for $g^2=g_0^2\left(1+{\sigma\over M_1}\right)$. 
Different points in the plots represent different values of $g$ and $M_1$.
The upper and lower dashed lines indicate the observational upper and lower limit 
of $f_{\rm NL}=2.5\pm 5.7$. We see that
to keep $f_{\rm NL}$ within the observational bound for arbitrary choice of $g$ we need $M_1\ge 2\times 10^{-3}$.
${\langle\sigma\rangle^*}$ has been chosen to be 0.2 to keep amplitude of iso-curvature perturbation under 
observational constraint. We see that for that for values of $M_1$ above the threshold mentioned, ${\cal P}_{\zeta}^{\rm MP}$ to ${\cal P}_{\zeta}$ remains under 0.15. Results are in $M_P=1$ unit.}
\label{fnl-con-linear}
\end{figure*}


 As done in subsection \ref{con_amp}, 
we take a set of 10000 random values of $g\over m$ and $M_2$ for 
the model in \Eqn{g-form-2}. The random values are generated with uniform distribution in 
the two dimensional parameter space of $g\over m$ and $M_2$ where the range for $g\over m$ is from 320 to 400 
and the range for $M_2$ is from $1\times 10^{-2}$ to $8\times 10^{-2}$. 
Then we fix ${\langle\sigma\rangle^*}$ to be 0.2. $N_g$ and $N_{gg}$ take different values
for different $g\over m$ as described in table \ref{table}.
So $f_{\rm NL}$ becomes function of only $g\over m$ and $M_2$. In \Fig{fnl-con-quad}
we see that for different values of $g\over m$, $f_{\rm NL}$ may behave arbitrarily, but 
there is a clear dependency of $f_{\rm NL}$ with $M_2$ as evident from \Eqn{fnl-M2}. This ensures that if we want to avoid fine tuning of
$g\over m$ and still want to keep $f_{\rm NL}$ within observational limit $2.5\pm 5.7$~\cite{Ade:2015ava},  
we need  $M_2\ge 4\times 10^{-2}$. This result immediately translates into the fact that
(see the right panel of \Fig{fnl-con-quad}) the amplitude of the power spectrum of
curvature perturbation produced via modulated preheating
will always remain within 13\% of that of the total curvature perturbations.

We repeat the same exercise for the model of \Eqn{g-form-1}.
Here $f_{\rm NL}$ takes the following form
\begin{eqnarray}\label{fnl-M1}
f_{\rm NL}^{\rm local}={5 {H^*}^4 N_g^2 g^2 \over 384 M_1 \pi ^4 P_{\zeta}^2 \left(M_1+{\langle\sigma\rangle^*} \right)}
\left(-\frac{N_{g} g}{4 M_1^2 \left(1+\frac{{\langle\sigma\rangle^*} }{M_1}\right)^{3/2}}+\frac{N_{gg} g^2}{4 M_1^2 \left(1+\frac{{\langle\sigma\rangle^*} }{M_1}\right)}\right)
\end{eqnarray}
We find that we need $M_1\ge 2\times 10^{-3}$ and 
in this case amplitude of the power spectrum of curvature perturbations produced via modulated preheating
remains below 15\% of that of the total curvature perturbations(see \Fig{fnl-con-linear}).   

{
We have discussed in the previous section, the iso-curvature perturbation may not show up in CMB if the universe has 
undergone an era of local thermal equilibrium. In that case, the constraint on the value of $\langle\sigma\rangle^*$
is not valid. Still we can draw some conclusions regarding the amount of curvature perturbations that can be allowed to be generated 
by modulated preheating. We see in \Fig{fnl-con-quad} and \Fig{fnl-con-linear} that the minimum acceptable values of $M_1$
and $M_2$ are much smaller than the fixed value of $\langle\sigma\rangle^*$. We have also understood from these figures
that when $M_1$ and $M_2$ tend towards $\langle\sigma\rangle^*$,
the ratio ${{{\cal P}_{\zeta}^{\rm MP}} / {\cal P}_{\zeta}}$ becomes smaller.}
Because of this monotonic nature, it can be safely said that the upper bound on ${\cal P}_{\zeta}^{\rm MP}/ {\cal P}_{\zeta}$
need not be revised even if we take much higher values of $M_1$ and $M_2$ than what have been shown in the figures. 

\begin{figure*}
\centerline{\epsfxsize=0.67\textwidth\epsfbox{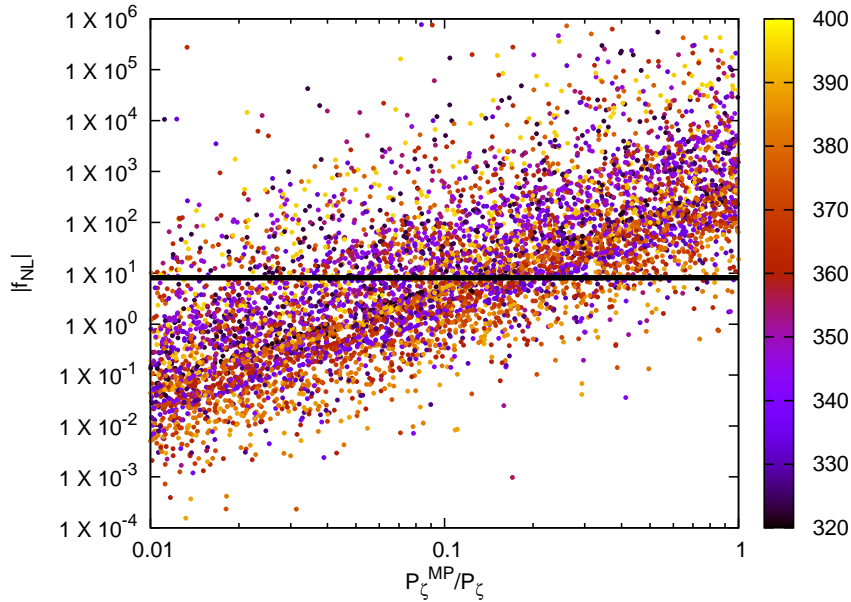}}
\caption{For $M_1$ and $M_2$  much smaller than $\langle\sigma\rangle^*$ the amount of curvature perturbation
generated by modulated preheating has to be less than 1\% of the total curvature perturbations to keep the absolute
value of $f_{\rm NL}^{\rm local}$ within the observational upper limit. 
The line shows the observational upper limit of $f_{\rm NL}^{\rm local} = 8.2$. Different colours 
represent different $g\over m$ values.}
\label{s-fnl}
\end{figure*}

{
Now let us investigate the case where $M_1$ and $M_2$ are much smaller than $\langle\sigma\rangle^*$. We notice from 
 \Fig{fnl-con-quad} and \Fig{fnl-con-linear} that the value of $f_{\rm NL}$ increases in this case.
In this limit \Eqn{fnl} can be written as
\begin{eqnarray}\label{fnl-ratio}
 f_{\rm NL}^{\rm local} 
\approx \frac{5 N_{gg}}{6 N_g^2}\left({{{\cal P}_{\zeta}^{\rm MP}}\over {\cal P}_{\zeta}}\right)^2\, .
\end{eqnarray}
In \Fig{s-fnl} we check how $f_{\rm NL}$ behaves with ${{{\cal P}_{\zeta}^{\rm MP}} / {\cal P}_{\zeta}}$ for different values of $g$.
}
Looking at the figure we see that the points in the plot predominantly appear in a band like region. For  ${{{\cal P}_{\zeta}^{\rm MP}} / {\cal P}_{\zeta}}<0.01$ this band lies within the allowed range of $|f_{\rm NL}|$. Thus the constraint in this case becomes tighter compared 
to earlier cases where the corresponding maximum value of ${{{\cal P}_{\zeta}^{\rm MP}} / {\cal P}_{\zeta}}$ was 0.13 and 0.15 respectively.
Combining all cases we now conclude that the maximum allowed value of  ${{{\cal P}_{\zeta}^{\rm MP}} / {\cal P}_{\zeta}}$ is 0.15.

\section{Conclusion}\label{results}
In this paper we have studied, both semi-analytically and numerically, the effects that can be 
produced in CMB if the coupling of the inflaton with another scalar field varies on super horizon scales.
Modulated reheating mechanism has been thought of producing total observed CMB perturbations. 
But here we have shown that its non-pertubative extension, modulated preheating, is not fit for that 
purpose. If modulated preheating has to be responsible for producing substantial large amount
of observed density perturbation in CMB and at the same time it has to keep the iso-curvature perturbations and 
local form non-gaussianity within observational limit, the parameters of the theory have to be unnaturally 
fine tuned. Only for certain lower bound on the mass scales related to the modulating field the theory 
becomes natural. But this lower bound on the mass scales immediately dictates that 
modulated preheating mechanism can at maximum be responsible for generating 15\% of the observed CMB fluctuations.

{
If the iso-curvature perturbations become adiabatic during the evolution of the universe,
the lower bound on the background value of the modulating field($\langle\sigma\rangle^*$) does not hold. 
Consequently the lower bounds on mass scales $M_1$ and $M_2$ can not be derived.   
But that could not relax the 
upper limit on the allowed amount of curvature perturbations produced by modulated preheating mechanism.
For any value of $\langle\sigma\rangle^*$ there are two possibilities. First is that the cutoff scales $M_1$ and $M_2$ are 
much smaller than $\langle\sigma\rangle^*$ and second possibility is the opposite of it. For the first category the
modulated preheating can generate 1\% of the observed CMB fluctuations at maximum if the 
absolute value of $f_{\rm NL}$ has to remain with observational limit, while for the second category it
can go up to around 13\% or 15\% depending on the model.
} 

From the point of view of model building the lower bound on mass scales $M_1$ and $M_2$ is
expected to be very useful. As we have discussed that in different models this scales come from different
physics. In string-related models these scales come from the compactification of some 
higher dimensions. In particle physics models these scales corresponds to the mass of intermediate
particles in the theory. Therefore constraints on $M_1$ and $M_2$ would indirectly provide bounds
on the physics behind producing modulated coupling constant. These constraints on $M_1$
and $M_2$ would be always applicable as long as  inflaton potential can be assumed to be like $m^2\phi^2$  
near its minima and inflaton's energy decays into another scalar field.

Moreover we have developed a semi-analytical method for calculating the order of the amplitude of the curvature
perturbation produced by this mechanism.  We have been able to supply the lattice results of 
the dependence of number of $e$-foldings during preheating era on the coupling constant of 
inflaton and a secondary scalar field. Therefore if there is a different model which has different dependence of coupling 
constant with the vev of a scalar field other than those discussed in our work, 
one can, in principle, use the lattice result to 
put a constraint on the cutoff scales for that framework.

\appendix
\section{Numerical details}\label{numerical}
Lattice simulation has been done using the publicly available code LATTICEEASY\cite{Felder:2000hq}.  We have taken a $32^3$ lattice 
which means a 3-dimensional lattice with 32 points along each edge. Length of the each side of lattice has 
been taken to be 20$H_i$, where $H_i$ is the Hubble parameter at the start of the simulation. For our case $H_i$ has 
the value of $4.9\times 10^{-7}$ and the value of $m$ has been fixed at $10^{-6}$. 
Time step in $m t$ was taken to be 0.001. Along with the double precession, this time step provides
an energy conservation accuracy(ECA) of the order of $10^{-4}$. Since the variation of $\delta N$ with $g$ is smooth and not spiky 
like the case with the variation of initial background value of $\chi$\cite{Bond:2009xx}, this ECA is sufficient to provide right accuracy for
the simulation.

Simulation terminates for each choice of $g\over m$ at a particular final Hubble parameter($H_e$) which has been taken to 
be $4.402\times 10^{-9}$. But it has been proved in \Eqn{deltaN2} that $\delta N$ is practically independent of the value of
$H_e$ as long as it denotes a Hubble parameter in radiation domination phase. Therefore different
choices of $H_e$ can lead to an overall change in $N_{\rm MP}$ but not in $\delta N_{\rm MP}, N_g$ or $N_{gg}$. So, the values obtained for the 
observable quantities are also independent of $H_e$.

\section*{Acknowledgement}
AM would like to thank Palash B Pal for various important discussions and suggestions. 
The authors acknowledge the Department of Atomic Energy (DAE, Govt. of India) for financial
assistance. 

\bibliographystyle{JHEP}
\bibliography{Modulated-Preheating-NG.bib}
\end{document}